\documentclass[11pt]{article}
 
\usepackage[utf8]{inputenc}
\usepackage[T1]{fontenc}
\IfFileExists{lmodern.sty}{\usepackage{lmodern}}{}
\usepackage[margin=1in]{geometry}
\usepackage{setspace}
\usepackage{amsmath,amssymb}
\usepackage{bm}
\usepackage{graphicx}
\usepackage{xcolor}
\usepackage[detect-all]{siunitx}
\usepackage{booktabs}
\usepackage{caption}
\usepackage[numbers,super,sort&compress]{natbib}
\usepackage{authblk}
\usepackage[hidelinks]{hyperref}
\usepackage{xurl}

\renewcommand{\figurename}{Fig.}
\renewcommand{\thefigure}{\arabic{figure}}   
 
\newif\ifshowptitles \showptitlesfalse      
\newcommand{\ptitle}[1]{\ifshowptitles\noindent\textcolor{blue}{\textbf{[#1]}}\par\fi}

\newcommand{\placeholderfig}[1]{%
  \fbox{\parbox[c][0.35\textheight][c]{0.95\linewidth}{\centering\large\textcolor{gray}{#1}}}}
\newcommand{\figorplaceholder}[2]{%
  \IfFileExists{figures/#1}{\includegraphics[width=\linewidth]{figures/#1}}{\placeholderfig{#2}}}
 
\title{\textbf{Agentic inverse design of \\ spatially programmed materials}}

\author[1,2,$\dagger$, *]{Denisse Córdova Carrizales}
\author[1,3,$\dagger$]{Abhijatmedhi Chotrattanapituk}
\author[1,2]{L. Dagny Davis}
\author[1,7]{Jeremy Ramsey}
\author[1,4]{Ryotaro Okabe}
\author[1,5,6]{Mouyang Cheng}
\author[1,2,5,*]{Mingda Li}

\affil[1]{Quantum Measurement Group, Massachusetts Institute of Technology (MIT), Cambridge, MA, USA}
\affil[2]{Department of Nuclear Science and Engineering, MIT, Cambridge, MA, USA}
\affil[3]{Department of Electrical Engineering and Computer Science, MIT, Cambridge, MA, USA}
\affil[4]{Department of Chemistry, MIT, Cambridge, MA, USA}
\affil[5]{Center for Computational Science and Engineering, MIT, Cambridge, MA, USA}
\affil[6]{Department of Materials Science and Engineering, MIT, Cambridge, MA, USA}
\affil[7]{Department of Applied Physics and Applied Mathematics, Columbia University, New York, New York, USA}

\begin{NoHyper}
\let\thefootnote\relax\footnotetext{$\dagger$ These authors contributed equally to this work.}
\let\thefootnote\relax\footnotetext{$\ast$ Corresponding authors: mingda@mit.edu, denisse@mit.edu}
\end{NoHyper}

\date{}

\makeatletter
\let\SIsavedauthor\@author
\makeatother
 
\begin{document}
\maketitle
 
\begin{abstract}
\noindent
Spatially programmed functional materials, from functionally graded alloys and metamaterials to quantum superlattices, use their spatial structure to tailor physical properties. However, translating a desired spatial structure into a practical fabrication recipe remains challenging due to the complex relationship between processing conditions and resulting material structures. Here we show that a large language model (LLM) agent can inverse-design such fabrication recipes by using a physics-based simulator of ion–matter interactions as a tool. Given an arbitrary depth profile of implanted ions or vacancies, the agent selects the ion species, energies, incident angles, and target material layers; writes, runs, and interprets simulations; and iteratively refines the recipe until the simulated profile matches the target, with minimal human input. Our work demonstrates physics-guided agentic inverse design as a practical approach to making the spatial structure of materials a programmable variable, opening new opportunities to explore emergent physical phenomena and functional phases of matter.

\end{abstract}

\newpage

\section*{Introduction}

Functionally graded materials \cite{miyamotoFunctionallyGradedMaterials1999a, woskoApplicationModelingFunctionally2005, chmielewskiMetalceramicFunctionallyGraded2016, landryQuantumTheoryFunctionally2026}, superlattices \cite{wanLayeredHybridSuperlattices2024, zhouModularSynthesisChemically2026, ivchenkoSuperlatticesOtherHeterostructures1997}, quasiperiodic stacks \cite{elhassouaniSurfaceElectromagneticWaves2006, kobialkaTopologicalSuperconductivityFibonacci2024, tokumotoSuperconductivityVanWaals2024a}, graded band gap heterostructures\cite{capassoBandGapEngineeringPhysics1987, lassalineGradientElectronicLandscapes2025}, and metamaterials \cite{jiaoMechanicalMetamaterials2023, qianGuidanceIntelligentMetamaterials2025} are all examples of spatially programmed functional materials. A programmed spatial structure adds a spatial degree of freedom to the tuning knobs commonly used to elicit coordinated macroscopic behavior from materials. These commonly used tuning knobs include applied electric fields, external pressure, and applied magnetic fields \cite{basovPropertiesDemandQuantum2017b, drozdovConventionalSuperconductivity2032015, liuTerahertzfieldinducedInsulatortometalTransition2012, zhongVanWaalsEngineering2017}. If we could inscribe an arbitrary spatial structure into a solid and theoretically describe its functionality \cite{landryQuantumTheoryFunctionally2026}, we could treat the spatial pattern of the material as a design variable and search for functionalities that emerge from the unique combination of the inscribed spatial design and physical composition of the material. To develop spatially programmed functional materials, the limiting step is increasingly the complex design of the fabrication process. Systematic exploration of spatially programmed functional materials thus requires inverting a target structure into material processing parameters.

\ptitle{Ion implantation can write any profile, and SRIM predicts it}
Spatially programmed functional materials can be synthesized via several techniques, including sputtering \cite{bhattacharjeeTopologicalAntiferromagneticVan2022}, molecular beam epitaxy \cite{panSuperconductivityQuintuplelayerSquareplanar2022}, and ion implantation \cite{mandal_tuning_2025, matsuiNonreciprocalCriticalCurrent2023}. Each method to create a spatially programmed material is typically a forward process: the user chooses the inputs, such as fluxes, temperatures, or beam parameters. To realize a superlattice profile using ion implantation, for example, the user must iterate over many ions, energies, angles, and superpose the resulting distributions by trial and error using an ion scattering simulation software such as SRIM/TRIM \cite{zieglerSRIMStoppingRange2010, paulCriticalOverviewRecent2013, kehayiasFittingAlgorithmOptimizing2021}.

The same bottleneck appears across materials design: theoretical and software forward models for growth, diffusion, and defect formation \cite{matssonIntroductionAnsysFluent2022, goodwinCanteraObjectorientedSoftware2021, pepperFiniteElementMethod2017, anderssonThermoCalcDICTRAComputational2002}, for example, are mature, but choosing the material processing parameters that produce a target structure heavily relies on human intuition. Each step of this design loop, such as choosing inputs, running a simulation, reading its output, and deciding what to change, is a decision followed by a software call, which can be performed by large language model (LLM) agents \cite{chowaLanguageActionReview2026, ferragLLMReasoningAutonomous2026, feng_agentic_2026}. LLM agents can even plan and execute chemical syntheses, run closed-loop materials discovery, and automate heuristic loops of discovery in physics \cite{nageleAgenticExplorationPhysics2026}. An agent that operates a physics simulator to inverse-design the experimental parameters that produce a selected spatial design in a material has yet to be reported to our knowledge.

\ptitle{What we did and found}
In this work, we show that an LLM agent inverse-designs spatially programmed materials by operating a physics simulator as a tool, and we demonstrate this on multi-step ion irradiation recipes for arbitrary depth profiles of both ion implants and vacancies. 
Given a target profile, the agent chooses the ion species, energies, fluences, incident angle, and layers of the target stack; runs SRIM/TRIM; then compares the simulated SRIM/TRIM profile with the target profile; and revises the recipe until the two agree with minimal human input. We apply the agent to target depth profiles of an irregular gradient, a superlattice gradient, and a Fibonacci superlattice. The agent also recommends the beam ramping rates and rest times to experimentally produce the target profiles.  We demonstrate agentic inverse design guided by a physics tool which turns the spatial structure of materials into a programmable variable through which to further explore emergent \cite{landryQuantumTheoryFunctionally2026} and functional phases of matter.

 
\section*{Results}
 
\subsection*{LLM agent closes design loop between target structure and fabrication process}

\ptitle{Architecture: coordinator, sub-agents and agent-built tool interfaces}
By connecting an LLM to physics tools, we build a system that executes material design tasks as shown in Fig. \ref{fig:architecture}a, b, c. To test whether an LLM agent can learn to operate SRIM without a human-designed
workflow, we gave the agent only short, high-level instructions and recorded the design
choices it made.

\subsection*{Agent reproduces irregular two-step gradient of hand-drawn sketch}
 
\begin{figure}[p]
  \centering
  \includegraphics[width=\textwidth]{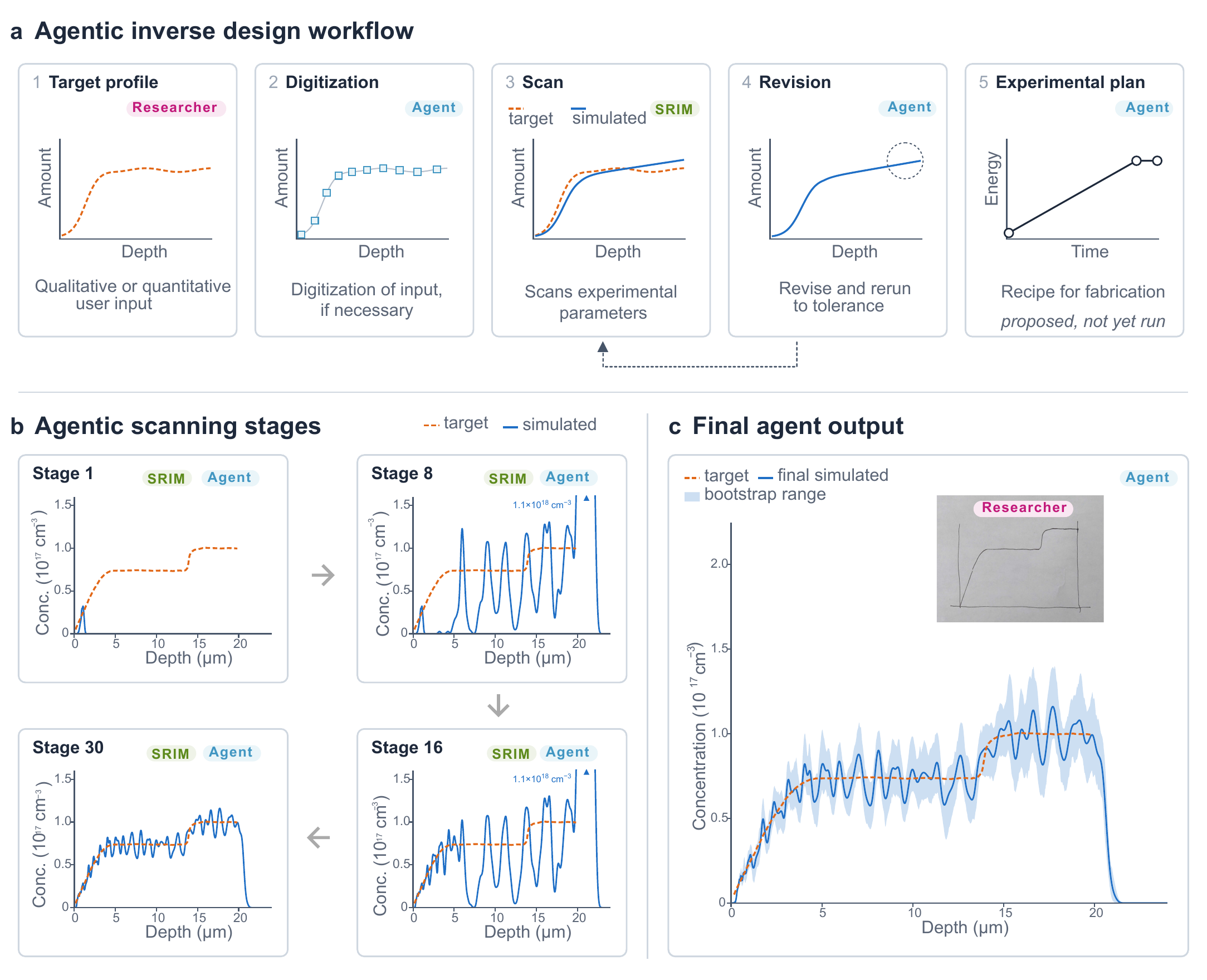}

  \caption{\textbf{Agent turns a hand-drawn sketch into a multi-energy implantation recipe}
  \textbf{(a)} Inverse-design workflow: (1)~the researcher gives a qualitative or quantitative target profile; (2)~the agent digitizes it, if necessary; (3)~the agent scans experimental parameters with SRIM; (4)~the agent revises the recipe and reruns it until the profile is within tolerance (dashed arrow); (5)~the agent outputs an ion implantation recipe for fabrication, which is proposed but not yet carried out.
  \textbf{(b)} Simulated profile (blue) compared with the target profile (orange dashed) after 1, 8, 16 and 30 beam energies are run by the agent. The model chooses to implant Bi into specified Si target.  These panels are retrospective cumulative refits: SRIM kernels were added in the order they were run, and all fluences were refit to the target at each stage. Only stages 23 and 30 reproduce optimizations the agent saved. The normalized root-mean-square error (NRMSE, Eq.~\ref{eq:nrmse}) is 78.3\%, 45.4\%, 36.5\% and 7.86\%, respectively. Triangles mark a peak of $1.1\times10^{18}$\,cm$^{-3}$ near \SI{21}{\micro\metre} that lies outside the plotted range.
  \textbf{(c)} Final design (blue) against the digitized target (orange dashed). The inset shows the researcher's hand-drawn sketch that served as the input. The recipe uses 28 energies with nonzero fluence, from 2.0 to \SI{195.8}{MeV}, for a total fluence of $1.55\times10^{14}$\,ions\,cm$^{-2}$. The NRMSE is 7.86\%. The shading shows a pointwise 95\% bootstrap interval from resampling SRIM ions at fixed dose, and it excludes uncertainty from fit selection and from the model.
  Ion and target: Bi ion into Si target. SRIM runs used 100 ions each. The ripples in the profile come from the fact that we force the model to create implanation peaks of approximately the same full width half max. Hence, the model needs to choose the combinations of energies and ion type that give roughly equal straggling through out the whole thickness. }
  \label{fig:gradient}
\end{figure}

\ptitle{From sketch to recipe}
We chose a gradient depth profile as a case study because functionally graded quantum materials are predicted to host emergent phenomena \cite{landryQuantumTheoryFunctionally2026}. 
To test whether the agent can work from loosely specified input, we drew the target by hand: a ramp from the surface to a plateau, then a step up to a second plateau, ending at \SI{20}{\micro\metre}. First, the agent digitizes the researcher's sketch, or takes a target defined numerically, as a concentration $C_\mathrm{target}(x)$ on a grid of depths (Fig.~\ref{fig:gradient}a). Then it fixes the ion, target, and angle of incidence. To find which energies place ions at the depths the target needs, the agent surveys how projected range varies with energy. It then runs SRIM at each selected energy to get a single-energy profile $k_i(x)$ per unit fluence (\ref{fig:gradient}b). Profiles from separate implants add linearly. Next, the agent evaluates the fitted design on three criteria: the residual (NRMSE, Eq.~\ref{eq:nrmse}), the spacing of projected ranges compared with the straggle, the overlap between adjacent implants. Where the residual shows an unfilled depth or an isolated peak, the agent adds or shifts energies and refits. The agent decided a 10\% NRMSE criterion on its own. Once the recipe is fixed, the agent checks it against independent SRIM runs. These runs use a different random seed and are not used in the fit. Finally, the agent exports the recipe, a list of energies and fluences (Fig. \ref{fig:gradient}c).

\ptitle{Error falls as energies fill in the depths the profile needs}
Figure~\ref{fig:gradient}b shows how the design converges as energies are added in the order the agent ran them, with all fluences refit at each stage. A single \SI{4.71}{MeV} implant covers only the first micron, and the NRMSE is 78.3\%. With eight energies, spread from 4.7 to \SI{225}{MeV}, the error falls to 45.4\%. The largest drops in error came when energies were added in the middle of the range. The 90.5, 123.4 and \SI{156.9}{MeV} implants filled the deep plateau and the step, and took the error from 28.9\% to 10.6\% across stages 19--22. Adding \SI{225}{MeV} had no effect, because that implant's range lies beyond the target, and the fit gave it zero fluence (stage~9). 

Importantly, the 30 stages in Fig.~\ref{fig:gradient}b are not 30 documented rounds of this loop. We reconstructed the progression retrospectively. We added the SRIM kernels in the order the agent ran them and refit all fluences to the target at each stage. Only stages~23 and~30 reproduce optimization checkpoints the agent saved during the campaign, and some earlier kernels were shared with later design goals in the same campaign. The progression therefore shows how the error falls as energies fill in the profile. It is not a record of individual decisions the agent made. 
The final recipe implants Bi into Si at 28 energies with nonzero fluence, from 2.0 to \SI{195.8}{MeV}, and reproduces the target with an NRMSE of 7.86\% (Fig. \ref{fig:gradient}c). 

The final profile oscillates around both plateaus with an amplitude of about $0.1$--$0.15\times10^{17}$\,cm$^{-3}$ (Fig.~\ref{fig:gradient}c). The agent chose $N = 100$ ions per SRIM run, trading statistical precision for speed. The 95\% bootstrap band is about as wide as the ripples of the different energy implants. The band, obtained by resampling the 100 ion depths at each energy with the fluences held fixed, reflects only SRIM counting statistics; the band excludes variation in the chosen recipe and systematic errors in SRIM’s stopping powers. Even with infinitely many ions though, adding finite separate peaks gives some ripple wherever the gap between neighboring peaks is larger than their width. 

\subsection*{Agent designs periodic and quasiperiodic superlattices}
 
\ptitle{Periodic superlattice: five energies, period within 1\%}
We next asked the agent to design ion implantation depth profiles for periodic and Fibonacci superlattices, which need sharp, well-separated peaks rather than smooth plateaus. Periodic \cite{lv_functional_2024} and Fibonacci superlattices \cite{de_dios-leyva_magneto-optical_2020} host functional properties for electromagnetic shielding, energy harvesting, optoelectronics, and thermoelectrics. 

\begin{figure}[p]
  \centering
  \includegraphics[width=\textwidth]{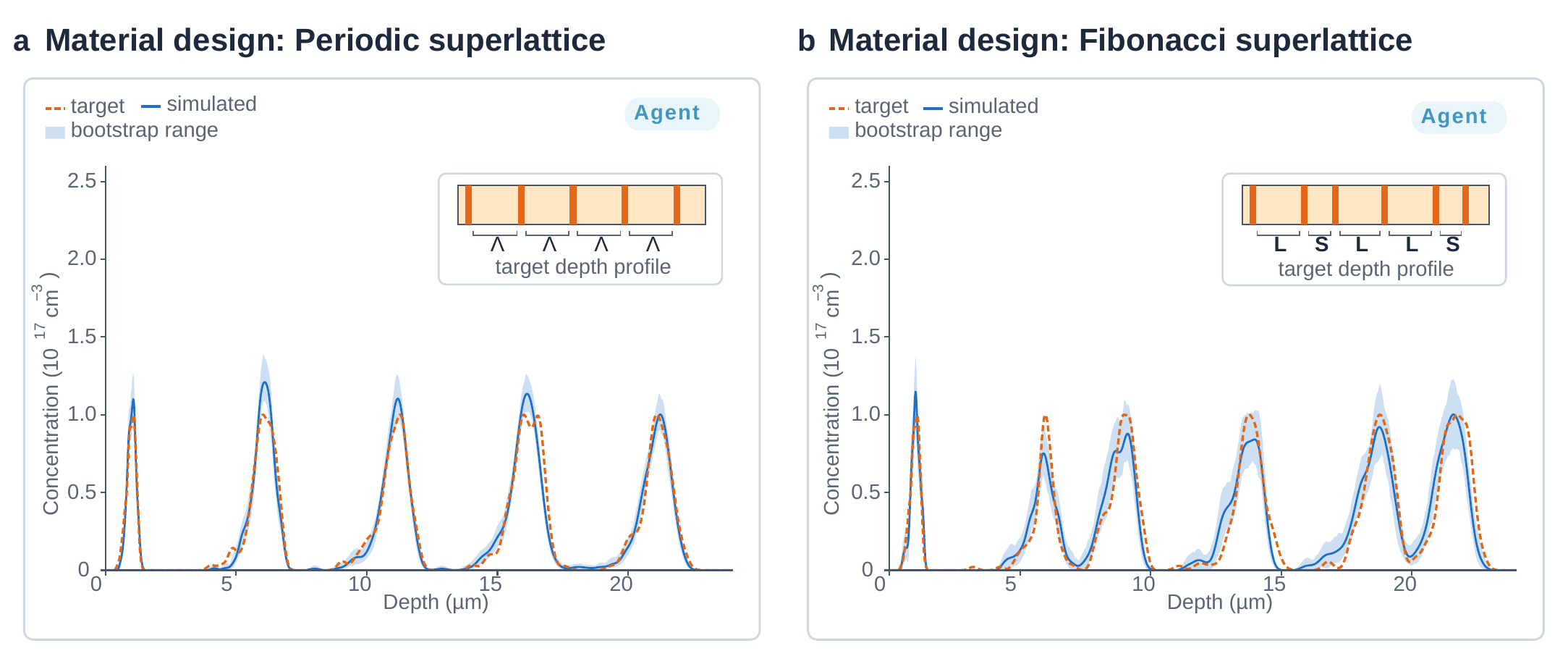}
  \caption{\textbf{Agent designs a periodic superlattice and a Fibonacci superlattice with implanted ions.} Bi ions implanted into Si target.
  \textbf{(a)} Periodic superlattice. The target (orange dashed) has five peaks at 1, 6, 11, 16 and \SI{21}{\micro\metre}, so the period is $\Lambda = \SI{5}{\micro\metre}$ (inset). The recipe uses five energies, from 4.71 to \SI{215.8}{MeV}, for a total fluence of $5.71\times10^{13}$\,ions\,cm$^{-2}$. An independent fixed-dose SRIM run (blue) gives peak spacings of 5.03, 5.10, 4.94 and \SI{5.10}{\micro\metre}, a peak-height coefficient of variation of 6.0\% and a worst valley-to-peak ratio of 1.2\%.
  \textbf{(b)} Fibonacci superlattice. The target profile follows the sequence L\,S\,L\,L\,S, with $L = \SI{4.854}{\micro\metre}$, $S = \SI{3.000}{\micro\metre}$, 
  and its peaks lie at 1.00, 5.85, 8.85, 13.71, 18.56 and \SI{21.56}{\micro\metre}. The recipe uses six energies, from 4.71 to \SI{225.2}{MeV}, for a total fluence of $6.64\times10^{13}$\,ions\,cm$^{-2}$. The independent confirmation run gives spacings of 4.90, 3.22, 4.86, 4.77 and \SI{2.84}{\micro\metre}, a peak-height coefficient of variation of 13.6\%, and a worst valley-to-peak ratio of 9.5\%.
  In both panels, the shading shows the pointwise 95\% fixed-dose bootstrap interval which excludes uncertainty from fit selection and from the model.}
  \label{fig:superlattice}
\end{figure}

A superlattice is a periodic stack of alternating layers of two or more materials (which are themselves periodic crystals), and the larger artificial periodicity can change the materials' electronic, phononic, or optical properties. For a periodic superlattice with period $\Lambda = \SI{5}{\micro\metre}$ and five peaks between 1 and \SI{21}{\micro\metre}, the agent proposed a sparse recipe of five energies from 4.71 to \SI{215.8}{MeV} (Fig.~\ref{fig:superlattice}a). An independent fixed-dose SRIM run of that recipe gives peak spacings of 5.03, 5.10, 4.94 and \SI{5.10}{\micro\metre}, with a mean of \SI{5.04}{\micro\metre} within 1\% of the target period. The peak heights are nearly uniform with a coefficient of variation of 6.0\%. The concentration between peaks falls to at most 1.2\% of the neighboring peak height, so the layers are well isolated. 
  
\ptitle{Fibonacci superlattice: quasiperiodic order preserved}
A Fibonacci superlattice is quasiperiodic: ordered but never repeating. It is built from two spacings, $L$ and $S$, arranged by the substitution rule\cite{merlin_quasiperiodic_1985}.
\begin{equation}
  L \rightarrow LS, \qquad S \rightarrow L, \qquad
  \frac{L}{S} \;=\; \varphi \;=\; \frac{1+\sqrt{5}}{2} \;\approx\; 1.618 .
  \label{eq:fibonacci}
\end{equation}
\noindent We asked for the sequence L\,S\,L\,L\,S and the agent decided on $L = \SI{4.854}{\micro\metre}$ and $S = \SI{3.000}{\micro\metre}$ (Fig.~\ref{fig:superlattice}b). The agent used six energies from 4.71 to \SI{225.2}{MeV}. The long spacings of 4.90, 4.86 and \SI{4.77}{\micro\metre} have a mean of \SI{4.84}{\micro\metre}, and the short spacings of 3.22 and \SI{2.84}{\micro\metre} have a mean of \SI{3.03}{\micro\metre}. The ratio of the means, $L/S = 1.60$, lies within 1.2\% of $\varphi$, so the quasiperiodic order is preserved.
 
\ptitle{Peak uniformity degrades where spacing approaches straggle}
The peak heights are less uniform, however. The coefficient of variation is 13.6\%, and the worst valley-to-peak ratio is 9.5\%. Both trends follow from the geometry. Straggle grows with depth, so the deep peaks broaden. The \SI{3}{\micro\metre} short spacings place neighboring peaks closer together than in the periodic case, so the peaks overlap and fill the valleys between them. This regime marks the resolution limit of ion implantation. When the layer spacing approaches the straggle, no choice of fluences restores full contrast. 
 
\ptitle{Sparse recipes suffice for peaks; gradients need dense ones}
Comparing Figs.~\ref{fig:gradient} and~\ref{fig:superlattice} shows that the size of the agent's recipe depends on the target. Five or six energies produce a superlattice because each peak corresponds to one implant. The two-step gradient needed 28 energies because a flat plateau requires overlapping implants. The agent chose both recipe sizes on its own. 

\subsection*{Agent designs defect profiles}
\begin{figure}[p]
  \centering
  \includegraphics[width=\textwidth]{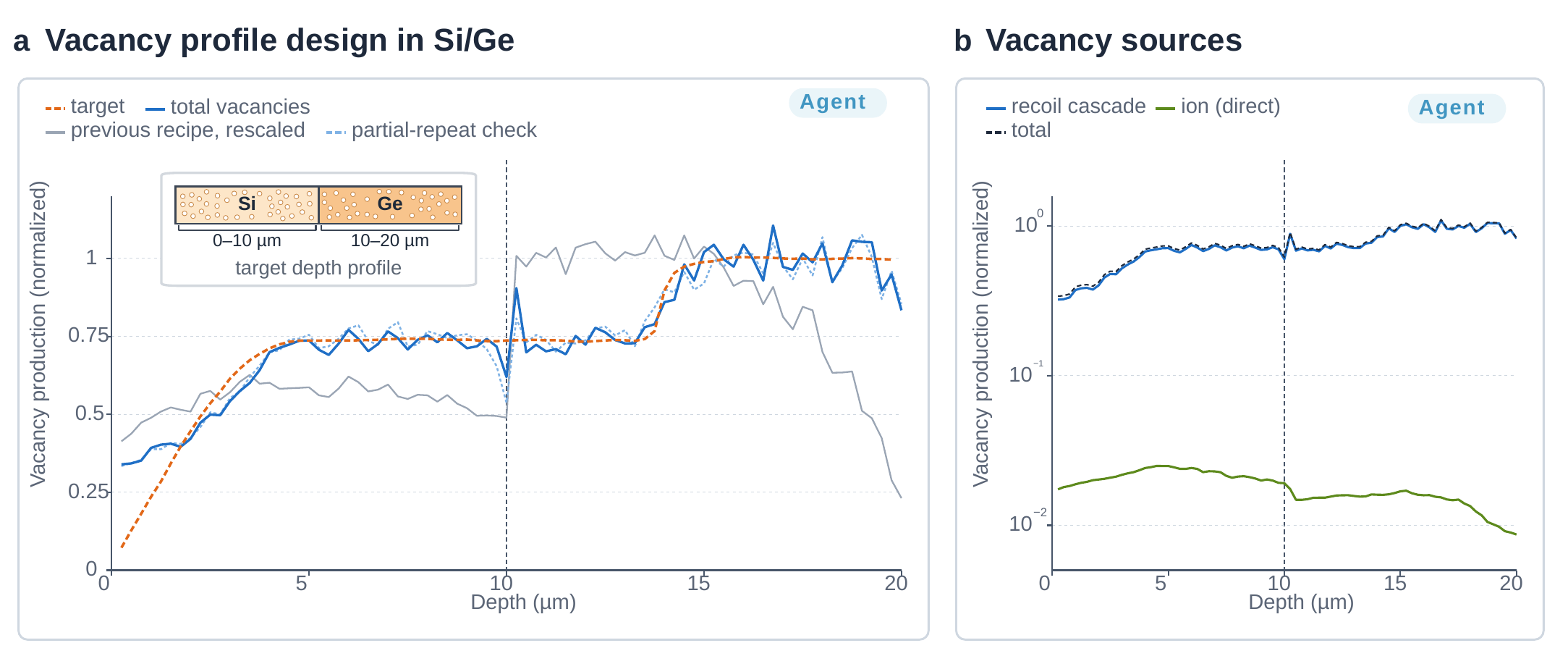}
\caption{\textbf{Agent designs gradient defect profile across a Si/Ge interface}
  \textbf{(a)} Normalized vacancy production versus depth for $^{209}$Bi at normal incidence into \SI{10}{\micro\metre} of Si on \SI{10}{\micro\metre} of Ge (dashed vertical line, interface). The target profile (orange dashed) is the hand-drawn sketch in Fig.\ref{fig:gradient}c, normalized so that the upper plateau equals 1. Whereas in Fig. \ref{fig:gradient} we asked the agent to create the gradient profile from ion implantation, here we ask the agent to create the gradient profile from vacancies induced by ion implantation in a two layer material stack. The agent's recipe (blue, total vacancies) uses 18 energies with nonzero fluence from 20 to \SI{260}{MeV}. These energies are chosen from 25 existing SRIM runs (previously run by the agent for Fig. \ref{fig:gradient}). The simulation achieves a normalized root-mean-square error (NRMSE, Eq.~\ref{eq:nrmse}) of 6.61\% over 0.25--\SI{19.75}{\micro\metre} from the target profile (79 bins; design criterion 10\% NRMSE) . The light-blue dotted curve tests the design against SRIM counting noise. Keeping the fitted fluences fixed, the agent replaces the kernels at five energies with independent SRIM runs (different seed, 250 ions). Only four of these energies have nonzero fluence, and together they carry 29.7\% of the total. The NRMSE changes from 6.61\% to 6.66\%. The design departs most from the target at the surface (0.25 µm): it gives 0.34 where the target is 0.07, a difference of 0.27 in units of the upper plateau. Every ion that reaches the deep plateaus also creates vacancies near the surface on its way in, and non-negative fluences cannot remove that damage. \textbf{(b)} The same design separated into vacancies from the recoil cascade (blue) and from the ion directly (green), with their sum (black dashed), on a logarithmic scale. Recoils produce 97.6\% of the vacancies across the fitted window. Vacancy production is normalized to the upper target plateau; the absolute level scales linearly with the total fluence. Vacancies are SRIM Kinchin--Pease estimates of primary damage, not surviving defects after recombination or annealing. Profiles use the native SRIM bins of \SI{250}{nm} without smoothing; the first \SI{250}{nm} is unresolved, and the bins at the interface average over both layers. No confidence band is shown because SRIM's aggregate vacancy tables do not support an event-level bootstrap.}
  \label{fig:vacancy}
\end{figure}

\ptitle{Same sketch, new target quantity: vacancy production}
For a broad range of materials, including ones that need to be radiation hard \cite{muhammad_radiationtolerant_2023} or ones that benefit from quantum defects \cite{bassett_quantum_2018}, the property of interest is the damage that ions leave behind. We therefore gave the agent the same hand-drawn sketch as in Fig.~\ref{fig:gradient} but asked it to match the depth profile of vacancy production in a two-layer target: \SI{10}{\micro\metre} of Si on \SI{10}{\micro\metre} of Ge irradiated with $^{209}$Bi at normal incidence (Fig.~\ref{fig:vacancy}a). Si/Ge heterostructures are used for microelectronics \cite{paul_sisige_2004}.

\ptitle{18 energies reproduce both plateaus with 6.6\% error}
The agent built the design from 25 SRIM runs it had already performed (for Fig. \ref{fig:gradient}) with no new simulations and found a recipe of 18 energies (from 20 to \SI{260}{MeV}) with nonzero fluence. The recipe reproduces the target with an NRMSE (Eq.~\ref{eq:nrmse}) of 6.61\%, below the 10\% design criterion, over the fitted window from 0.25 to \SI{19.75}{\micro\metre}.  The energy implants which carry the largest share of the dose are \SI{260}{MeV} which takes 18.0\% of the total fluence and \SI{64.7}{MeV} takes 15.3\%.

\ptitle{Recoil cascades produce 98\% of the vacancies}
Separating the two sources of damage shows why the design hinges on the recoil cascade (Fig.~\ref{fig:vacancy}b). Across the fitted window, the recoil cascade produces 97.6\% of the vacancies, and the ion itself only 2.4\%. The vacancy profile is thus set almost entirely by the energy that the Bi ions pass to Si and Ge recoils, and the ion contribution is a small correction.

 
\ptitle{Remaining error: the surface, the interface and statistics}
The largest error lies at the surface. No non-negative recipe can reach such a low surface value because every implant deep enough to fill the plateaus also damages the surface on its way in. Consistent with this, the fit assigned zero fluence to all six energies below \SI{20}{MeV}, from 2.0 to \SI{16.9}{MeV}, whose damage would be confined to the near-surface region. A second artifact sits at the interface because SRIM averages the vacancy production of a bin that straddles both layers. 

As an additional design case study, we asked the agent to create the same gradient design on a Si/Ge heterostructure but now with ion concentration instead of vacancies. The results are shown in Supplemental Fig. \ref{supplement Si Ge heterostructure ion implants}.

\subsection*{Agent designs within limits of real accelerator and returns beam schedule}

\begin{figure}[p]
  \centering
  \includegraphics[width = \textwidth]{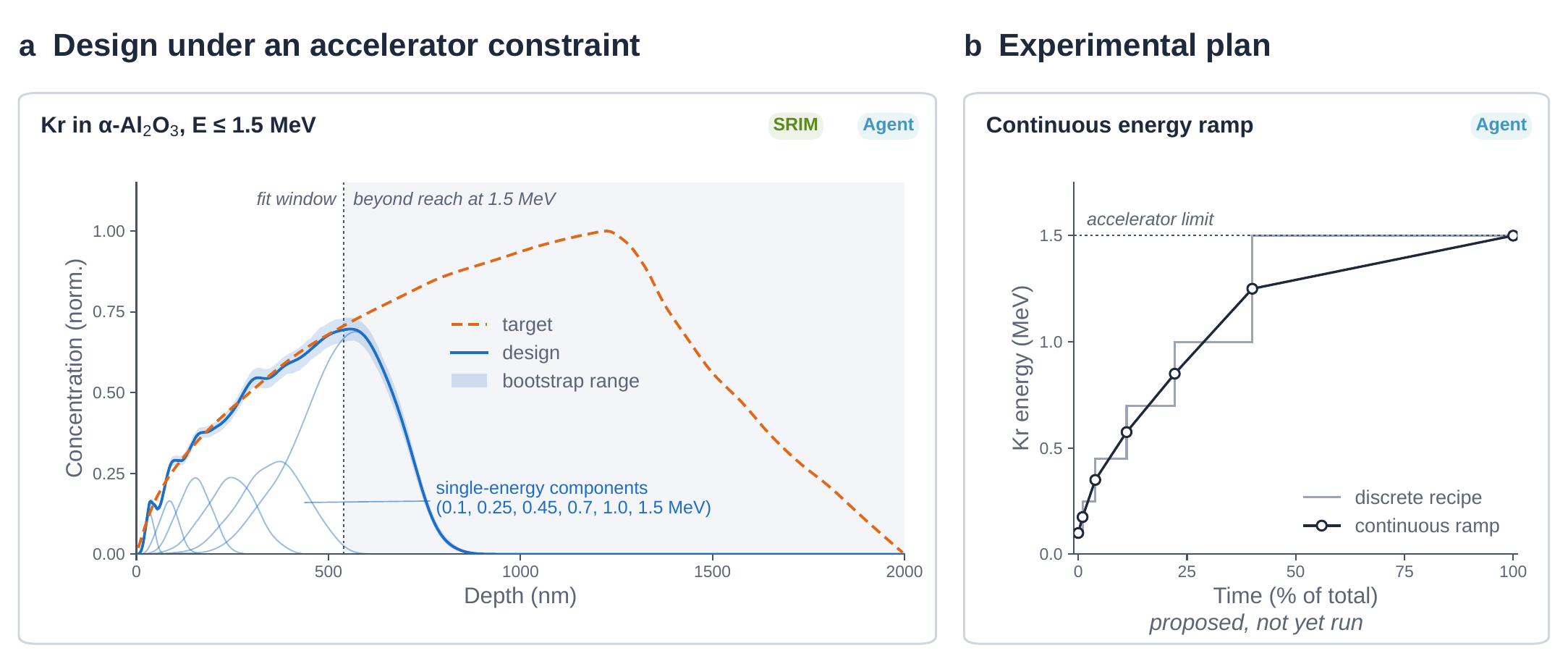}
  \caption{\textbf{Under a maximum beam energy of 1.5\,MeV, the agent fits the depths the beam can reach and converts the recipe into a continuous energy ramp.}
  \textbf{a},~Kr implantation in a \SI{2}{\micro\metre} film of $\alpha$-Al$_2$O$_3$. The target (orange dashed) is normalized to its maximum near \SI{1.2}{\micro\metre}. The design (blue) uses six energies, 0.10, 0.25, 0.45, 0.70, 1.0 and \SI{1.5}{MeV}, with 1.0, 2.9, 7.2, 11.1, 17.8 and 60.0\% of the total fluence; thin lines show their individual contributions. The fit window (white) ends at \SI{540}{nm}, and the mean Kr depth is at \SI{1.5}{MeV}; depths beyond it (grey) cannot be reached at the allowed energies. The normalized root-mean-square error (NRMSE, Eq.~\ref{eq:nrmse}) is 4.57\% over the fit window (54 bins of \SI{10}{nm}) and 57.3\% over the full film. Profiles are Gaussian kernel density estimates of the SRIM stopping depths, with a bandwidth of 0.30 times the standard deviation of each energy's depths. The shading is a pointwise 95\% bootstrap range from 500 resamplings of the stopping depths at fixed fluence and bandwidth, and it excludes uncertainty from the fit and from the model. SRIM runs used 2000 ions each, at normal incidence, in the Kinchin--Pease (``Quick'') mode of SRIM-2013. 
  \textbf{b},~Energy--time plan for the recipe in \textbf{a}. The discrete recipe (grey) holds each energy for its fraction of the total fluence. The continuous ramp (black; Eqs.~\ref{eq:knots}--\ref{eq:ramp}) passes through knots (circles) at the midpoints between neighboring energies, 100, 175, 350, 575, 850, 1250 and \SI{1500}{keV}, at 0, 1.0, 3.9, 11.1, 22.2, 40.0 and 100\% of the beam time, assuming constant ion flux. The dotted line marks the \SI{1.5}{MeV} limit. The ramp is proposed and has not yet been simulated or carried out.}
  \label{fig:constraint}
\end{figure}

\ptitle{A target deeper than the beam can reach}
The designs in Figs.~\ref{fig:gradient}--\ref{fig:vacancy} assumed that any beam energy was available. In practice, an accelerator sets a maximum energy, and that limit sets the deepest layer the profile can reach. To test whether the agent respects such a limit, we asked it to match a sketched Kr profile in a \SI{2}{\micro\metre} film of $\alpha$-Al$_2$O$_3$ ($\rho = \SI{3.98}{g\,cm^{-3}}$) with beam energies no higher than \SI{1.5}{MeV} (Fig.~\ref{fig:constraint}a). The target rises from the surface to a maximum near \SI{1.2}{\micro\metre} and then falls to zero at \SI{2}{\micro\metre}. SRIM gives a mean Kr depth of only \SI{545}{nm} at \SI{1.5}{MeV}, so most of the target lies beyond the reach of the allowed beam.

\ptitle{The agent fits only the reachable depths}
Rather than spreading the error over the whole film, the agent restricted the fit to the depths the beam can reach, 0--\SI{540}{nm}, set by the mean depth at the maximum energy. Within this window, it chose six energies, 0.10, 0.25, 0.45, 0.70, 1.0 and \SI{1.5}{MeV}, with mean depths from 37 to \SI{545}{nm}, and fit their fluences to the target (Eq.~\ref{eq:superposition}). The \SI{1.5}{MeV} implant carries 60.0\% of the fluence and the \SI{0.1}{MeV} implant 1.0\%. The design reproduces the target with an NRMSE (Eq.~\ref{eq:nrmse}) of 4.57\% over the fit window, normalized to the global maximum of the target, or 6.49\% when normalized to the target maximum within the window. 

\ptitle{A smooth estimator and a sampling band}
Each SRIM run here tracked 2000 ions, so we estimated each single-energy profile with a Gaussian kernel density estimate (KDE) of the stopping depths, with a bandwidth of 0.30 times the standard deviation of those depths (3--\SI{35}{nm}). To bound the sampling error, the agent resampled the stopping depths of each energy 500 times with fixed fluences and bandwidths and took the pointwise 2.5th and 97.5th percentiles. The bootstrap band excludes uncertainty from the fit and from the SRIM model.

\ptitle{From discrete energies to a continuous ramp}
A recipe of discrete energies requires the operator to stop the beam and retune it between implants. In the last step of the workflow (Fig.~\ref{fig:gradient}a, step~5), the agent instead converted the recipe into a single continuous energy sweep at constant ion flux (Fig.~\ref{fig:constraint}b). It divided the energy axis at the midpoints between neighboring energies, which gives boundaries $e_0, \dots, e_6$ of 100, 175, 350, 575, 850, 1250 and \SI{1500}{keV}, and assigned each interval the time fraction $w_k$ of the discrete energy it contains. The knot times are the cumulative fractions,
\begin{equation}
  t_k \;=\; \sum_{j=1}^{k} w_j, \qquad t_0 = 0, \quad t_6 = 1,
  \label{eq:knots}
\end{equation}
\noindent and the energy rises linearly between knots,
\begin{equation}
  E(t) \;=\; e_{k-1} + \left(e_k - e_{k-1}\right)\frac{t - t_{k-1}}{t_k - t_{k-1}}, \qquad t_{k-1} \le t \le t_k .
  \label{eq:ramp}
\end{equation}
\noindent At constant flux, the fluence in each interval is unchanged, and it is spread evenly over that interval's energies,
\begin{equation}
  \frac{d\Phi}{dE} \;=\; \frac{w_k\,\Phi_\mathrm{tot}}{e_k - e_{k-1}}, \qquad e_{k-1} \le E \le e_k ,
  \label{eq:ramp_fluence}
\end{equation}
\noindent where $\Phi_\mathrm{total}$ is the total fluence. The resulting ramp climbs steeply through the low energies, reaching \SI{850}{keV} after 22\% of the beam time, and spends the last 60\% sweeping from 1250 to \SI{1500}{keV}. We have not yet simulated nor experimentally verified the implanted profile of the ramp itself. The ramp is therefore a starting plan for an operator rather than a validated recipe.


\subsection*{Agent respects physical limits despite user request}

\begin{figure}[p]
  \centering
  \includegraphics[width = \textwidth]{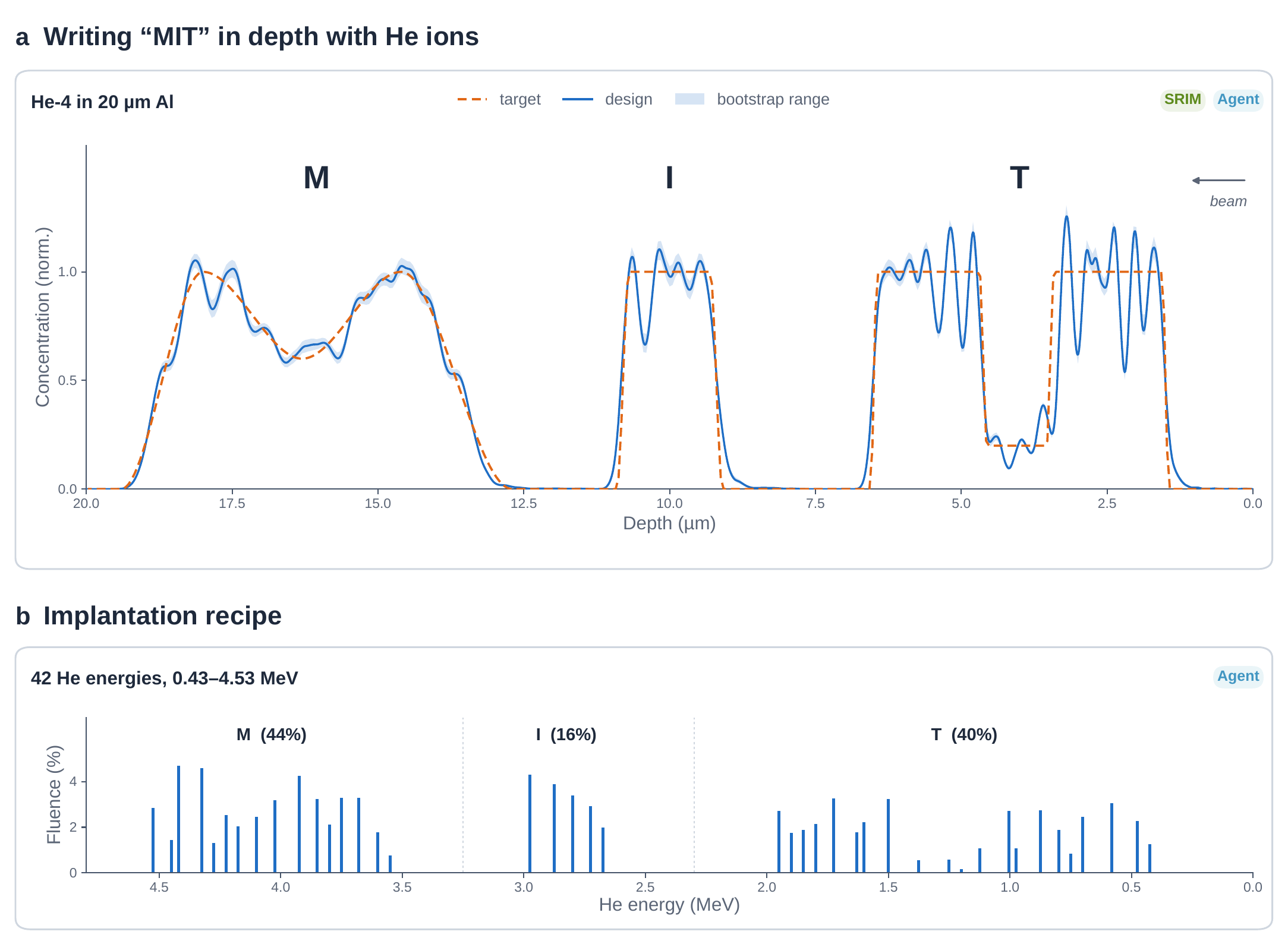}
  \caption{\textbf{The agent writes the letters ``MIT'' into the depth profile of He in Al, reproducing smooth shapes closely and sharp walls only down to the ion straggle.}
  \textbf{a},~He concentration versus depth in a \SI{20}{\micro\metre} Al film ($\rho = \SI{2.70}{g\,cm^{-3}}$) at normal incidence, normalized to the design peak. Depth increases from right to left, so the beam (arrow) enters at the right and the profile reads ``MIT'' on the back of the sample. At our request, the agent reversed the letter order along depth to T-I-M because straggling is smaller at shallower depths. The target (orange dashed) has 100\,nm walls on the T and 200\,nm walls on the I. The design (blue) is the recipe in \textbf{b}, shown as Gaussian kernel density estimates of the SRIM stopping depths, with a bandwidth of 0.30 times the standard deviation of each energy's depths. The shading is a pointwise 95\% bootstrap range from 500 resamplings of the stopping depths at fixed fluence and bandwidth, and it excludes uncertainty from the fit, the target and the model. The normalized root-mean-square error (NRMSE, Eq.~\ref{eq:nrmse}) is 12.8\% over 400 bins of \SI{50}{nm} and 6.7\% for the same profile in 200\,nm bins. The 10--90\% widths of the walls are 0.22 and \SI{0.18}{\micro\metre} for the T and 0.33 and \SI{0.18}{\micro\metre} for the I (entry and exit).
  \textbf{b},~Fraction of the total fluence at each of the 42 He energies, from 0.425 to \SI{4.525}{MeV}, grouped by the letter each writes and plotted with energy increasing to the left to match \textbf{a}; percentages give each letter's share. The fluences were fit by non-negative least squares to the stopping positions binned at \SI{50}{nm}, with the M energies ($E \ge \SI{3.5}{MeV}$) held at an earlier recipe. For a peak concentration $C_0 = 10^{18}$\,cm$^{-3}$, the total fluence is $9.86\times10^{14}$\,ions\,cm$^{-2}$.
  SRIM runs used 2000 ions each in the Kinchin--Pease (``Quick'') mode; 45 were reused from earlier designs and two (2.675 and \SI{3.0}{MeV}) were new. The design neglects channeling, He diffusion and retention, and sputtering.}
  \label{fig:mit}
\end{figure}

\ptitle{A target with sharp walls, plateaus and smooth curves}
To test the agent on a depth profile that is not physically possible, we asked it to write the letters ``MIT'' into a \SI{20}{\micro\metre} Al film ($\rho = \SI{2.70}{g\,cm^{-3}}$, normal incidence) using He ions. The results are shown in Fig.~\ref{fig:mit}. 
The agent reused 45 SRIM runs from earlier designs and added two energies, 2.675 and \SI{3.0}{MeV}, to cover the walls of the letter I. Each run tracked 2000 He ions. Rather than upsampling coarse SRIM profiles, the agent binned the individual stopping positions at \SI{50}{nm} and fit the fluences by non-negative least squares over all 400 bins (Eq.~\ref{eq:superposition}). It held the fluences of the M energies ($E \ge \SI{3.5}{MeV}$) fixed at an earlier recipe and refit only the shallower energies. The final recipe uses 42 energies with nonzero fluence from 0.425 to \SI{4.525}{MeV} (Fig.~\ref{fig:mit}b). The letters separate cleanly in energy: 21 energies below \SI{2}{MeV} write the T with 40\% of the fluence, five energies from 2.675 to \SI{2.975}{MeV} write the I with 16\%, and 16 energies above \SI{3.5}{MeV} write the M with 44\%. As in Fig.~\ref{fig:vacancy}, the design is normalized, so it scales to any peak concentration $C_0$. The agent design reproduces the M and I letters legibly but fails to write the letter T. 
\section*{Discussion}
Our results show that a general-purpose LLM agent can close the loop between a materials design idea and a fabrication recipe by using an ion-matter simulator as a tool. The agentic material design loop demonstrated here extends beyond ion implantation to any fabrication process with an established physics model. By connecting to mature forward models for thin-film growth, diffusion, reactive flow, or phase equilibria, an LLM agent could translate a target structure into sputtering schedules, growth sequences, or anneal profiles. Feeding experimental data back into the loop, such as secondary ion mass spectrometry depth profiles in the case of this work, would let the agent refine its designs against real data---a move toward autonomous laboratories.

The ion implantation design in Fig. \ref{fig:mit} begs the questions: what other spatial designs can we write into materials, and what fundamental questions and practical utilities do they enable? The letters in Fig. \ref{fig:mit} suggest a first possibility with many other possibilities. For instance, a pattern written a few microns below the surface of a material is invisible from outside, so letters, a barcode, or a random sequence could serve as a tamper-evident, physically unclonable label for chips and components \cite{gaoPhysicalUnclonableFunctions2020,arppePhysicalUnclonableFunctions2017}. An asymmetric depth profile of dopants could direct the flow of particles in devices \cite{hanggiArtificialBrownianMotors2009, villegasSuperconductingReversibleRectifier2003}. A composition profile that sweeps steadily through depth could place an entire phase diagram inside one sample \cite{nakamuraTopologicalTransitionBi2011a}. Aperiodic sequences could be studied for optoelectronics \cite{jagannathanFibonacciQuasicrystalCase2021, vardenyOpticsPhotonicQuasicrystals2013} and used to explore topological states and adiabatic light pumping \cite{krausTopologicalStatesAdiabatic2012}. Finally, a carefully designed spatial profile could be used to more effectively simulate neutron damage \cite{zinkleOpportunitiesLimitationsIon2018} or study defect interface interactions \cite{beyerleinDefectinterfaceInteractions2015}.

Because our agent can attempt to reproduce even unconventional shapes with ion-matter simulations, the materials design question shifts from how to fabricate a given structure toward which physically motivated structures are worth creating. With our agentic framework, inverse design experiments that probe which spatial arrangement of a material maximizes a given property become more tractable. Spatially programmed materials thereby become a platform for the rational and predictive design of emergent and functional phases of matter.

\section*{Methods}
 
\subsection*{SRIM simulations}
SRIM/TRIM (Stopping Range in Materials), first introduced in 1985, is a computer program that computes ion stopping ranges, damage, and other kinetic phenomena in target materials by unifying experimental data and theoretical frameworks \cite{zieglerSRIMStoppingRange2010}. SRIM/TRIM is a binary collision approximation Monte Carlo code which considers the collision of only two atoms at a time and chooses stochastic impact parameters to generate each collision in the amorphous targets. To efficiently keep track of the paths of large numbers of ions, SRIM/TRIM makes several physical approximations like using an analytic formula to determine interatomic collisions and assuming a free flight path between collisions. Each path begins with an ion energy, position, and incident angle. Input parameters of the software are shown in Supplemental Fig. \ref{supplement SRIM workflow}. SRIM/TRIM can handle incident ion (Z = 1-92) energies from 10 eV to 2 GeV/amu with compound target materials of up to eight layers. Importantly, SRIM/TRIM does not evaluate thermal effects, assuming that the simulated target is at 0 K. 

\begin{figure}[p]
  \centering
  \includegraphics[width=\textwidth]{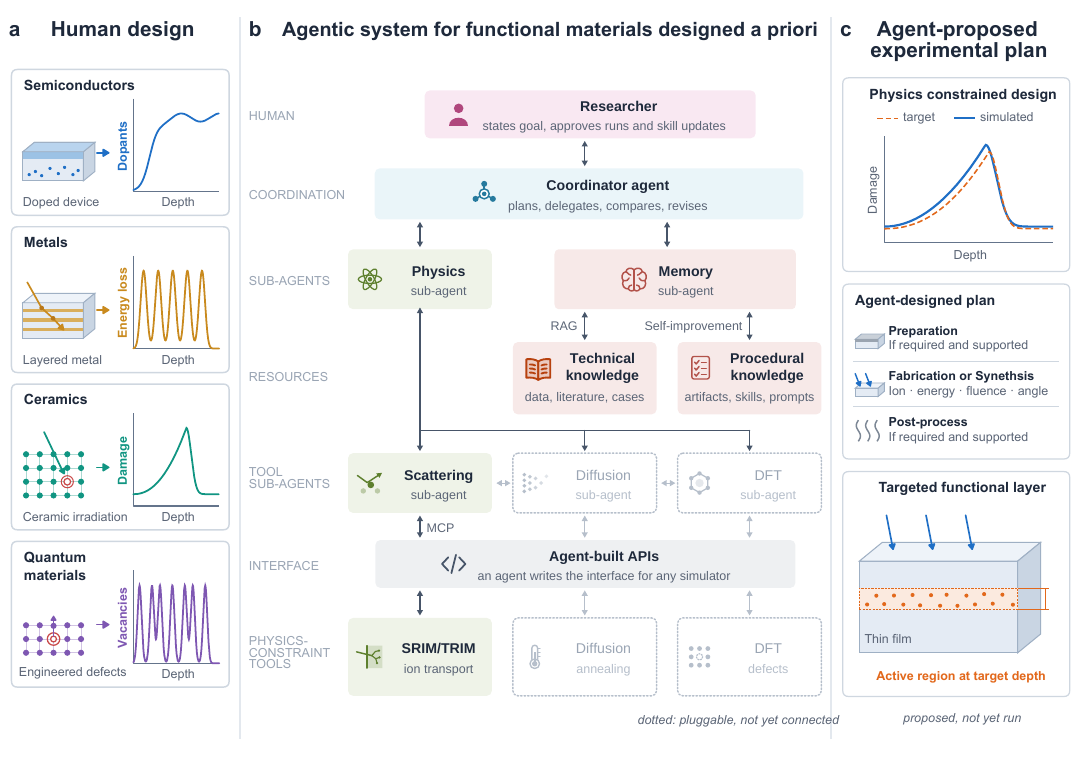} 
  \caption{\textbf{An agentic system designs the processing route for a functional material from a target spatial profile.}
  \textbf{(\textbf{a})} Selection of functional materials whose properties depend on a designed depth profile, such as dopant gradients in semiconductor devices, energy-loss modulation in layered metals, damage profiles in irradiated ceramics and engineered vacancy superlattices in quantum materials. Today a person typically designs each of these through either experimental or computational trial and error.
  \textbf{(\textbf{b})} Architecture of agentic inverse design system for spatial profiles in functional materials. A human states the profile goal (either quantitatively or qualitatively) and approves simulator runs and skill updates. A coordinator agent plans, delegates, compares and revises, supported by physics and memory sub-agents. The memory sub-agent accesses literature, past cases, relevant files through retrieval augmentation generation (RAG) and stores memory, skills and prompts for self-improvement. Tool sub-agents use Model Context Protocol (MCP) servers to connect with agent-built application programming interfaces (APIs) of physics simulators. In this work, the scattering sub-agent runs SRIM/TRIM for ion scattering in materials. The dotted gray lines for the diffusion (annealing) and density functional theory (DFT) (defects) modules emphasize additional sub-agents and tools which can be accessed to coordinate material design. Diffusion and DFT tools have not yet been connected in this agentic system. 
  \textbf{(\textbf{c})} The agent's output is an experimental plan. The output includes a physics-constrained design of target material profiles (illustrated here for damage) and a recipe (ion, energy, fluence and angle in the case of ion implantation). Note that the damage profile is illustrative, not computed. Also note that our current tool can only propose irradiation steps. The preparation and post-processing steps are shown to indicate the potential of connecting additional physics tools.}
  \label{fig:architecture}
\end{figure}

First, we supplied the agent with the SRIM documentation
\cite{zieglerSRIMStoppingRange2010} and asked it to build a Python application programming interface (API) for SRIM. The agent built the API with human input, and the full API design process is detailed in Supplemental Fig. \ref{supplement API and GUI}. Then, we asked the agent to propose a Model Context Protocol (MCP) server for this API which we reviewed and approved (Fig. \ref{fig:architecture}b).  Next, we briefly stated our project goal to inverse design ion implantation recipes to reproduce user-provided target depth profiles. We asked the agent create a virtual environment, connect to SRIM via the Python API, create two subagents to manage its workflow, and set up self-improvement protocols. The self-improvement protocols are detailed in Supplemental Fig. \ref{supplemental self improvement}. Supplemental Fig. \ref{supplement procedural memory} further details how the agent was instructed to update its workflows.

\ptitle{SRIM as the first connected simulator; human approval of runs}
In this work, the scattering sub-agent drives SRIM/TRIM, which calculates how ions travel through matter\cite{zieglerSRIMStoppingRange2010}. We input each design goal and every update to the agent's skills, so a person stays responsible for the physical decisions. The system outputs an experimental plan to recreate the target depth profile with ion implantation (Fig.~\ref{fig:architecture}c), detailing which ion, energies, fluences, angle to use, and target material stacks to use. The experimental output plan depends on the experimental constraints we detail in the input prompt.
 
\ptitle{The superposition model and error metric the agent optimizes}
The combined concentration of ions implanted is the fluence-weighted sum of single-energy SRIM profiles,
\begin{equation}
  C_\mathrm{sim}(x) \;=\; \sum_{i=1}^{n} \Phi_i \, k(x; E_i),
  \label{eq:superposition}
\end{equation}
\noindent where $x$ is depth, $\Phi_i$ is the fluence of implant $i$ (ions  cm$^{-2}$), and $g(x;E_i)$ is the SRIM ion distribution per unit fluence at energy $E_i$ (cm$^{-1}$). The total fluence is defined as 
\begin{equation}
    \Phi_{total}= \sum_i \Phi_i.
\end{equation}

The agent chooses the energies $\{E_i\}$ and fluences $\{\Phi_i \ge 0\}$ that minimize the normalized root-mean-square error (NRMSE) over the design depth window,
\begin{equation}
  \varepsilon \;=\; \frac{1}{\max_x C_\mathrm{target}(x)}
  \left[ \frac{1}{N_x} \sum_{j=1}^{N_x}
  \big( C_\mathrm{sim}(x_j) - C_\mathrm{target}(x_j) \big)^2 \right]^{1/2}.
  \label{eq:nrmse}
\end{equation}
\noindent 

\ptitle{Vacancy kernels are tracks, not peaks}
For vacancies, the forward model is again a superposition of single-energy SRIM runs, now of vacancy production per ion, $v(x;E_i)$ (vacancies ion$^{-1}$\,cm$^{-1}$),
\begin{equation}
  V(x) \;=\; \sum_{i=1}^{n} \Phi_i \, v(x;E_i), \qquad \Phi_i \;=\; C_0\, a_i ,
  \label{eq:vacancy_superposition}
\end{equation}
\noindent where $\Phi_i$ is the fluence at energy $E_i$ (ions\,cm$^{-2}$), $C_0$ is the vacancy density chosen for the upper plateau (vacancies\,cm$^{-3}$), and $a_i$ (ion\,cm\,vacancy$^{-1}$) are the coefficients the agent fits. Each $v(x;E_i)$ sums the vacancies made directly by the ion and those made by the recoil cascade. SRIM estimates them in the Kinchin--Pease approximation. Unlike an ion-concentration kernel, $v(x;E_i)$ is not localized near the projected range. Every ion produces vacancies along its whole track, with a maximum near the end of range. A deep implant therefore also damages every shallower layer, and the fitting problem is far less local than in Figs.~\ref{fig:gradient} and~\ref{fig:superlattice}.

The designs in this work inherit the approximations of SRIM: an amorphous target at 0~K, no channeling, no change in composition or sputtering at high fluence, and no defect recombination or diffusion \cite{zieglerSRIMStoppingRange2010}. SRIM stopping powers agree with experimental data to 11.4\% on average, from 6.6\% for heavy ions to 19.2\% for H \cite{bivorthaiekESPNNNovelElectronic2022}, so the depths of the designed features likely carry a systematic uncertainty of similar order. The vacancy designs further rely on the Kinchin--Pease estimate, whose absolute vacancy counts differ from full-cascade calculations \cite{zieglerSRIMStoppingRange2010}. We therefore normalized these designs and treat their shape, not their magnitude, as the result. Two limits apply to any recipe, independent of the agent. 

Though we generated the profiles of ion range distributions and displacements per atom (both in one dimension) in this work through the Python API connection, the agent is able able to interact directly with the SRIM GUI albeit much more slowly than through the Python API.  All simulations in this work were run using Kinchin-Pease Quick Mode in SRIM. 

Though we ran the SRIM simulations in this work on a Windows OS computer, we were also able to connect an LLM agent to SRIM on a Mac OS computers with the instructions in 
\href{https://github.com/antonio-padalino/SRIM-on-Mac}{this GitHub repository.}

\subsection*{Agent and tools}
We used GPT-6.1 Sol and all sub-agents are the same model but separate instances (clones of each other). We used the following prompt to generate the inverse designs of ion implantation depth profiles for Fig. \ref{fig:gradient} and Fig. \ref{fig:superlattice}.

\begin{quote}
    \textit{Given a silicon target, choose an appropriate ion type to perform the experiment to come up with the doping recipe for the following cases:}

    \textit{1. With supplied figure of a hand drawn depth profile, find the recipe of multiple energy implantations that give the depth profile close to the image. You can choose the height scale of the sketch.}

    \textit{2. Using the same ion, come up with the recipe for pseudo lattice of implants. Focus your optimization on the equal separation of peaks and distinctiveness.}

    \textit{3. Perform similar task to 2, but for Fibonacci lattice.}
\end{quote}

The workflow of the memory sub-agent in Fig. \ref{fig:architecture}b is detailed in Supplemental Fig. \ref{supplement RAG}, and details of the SRIM dataset available to the agent are shown in Supplemental Fig.  \ref{supplement 20000 database}. Performance of the agent with considering reuse evidence and external RAG is shown in Supplemental Fig. \ref{supplement fig:evaluation}. 

\subsection*{Python}
Our agentic system is compatible with Python 3.13+ and requires Python packages  mcp (version 2.3.0) and jsonschema (version 4.26.0). 
 

\section*{Code availability}
Python APIs and local MCP tools for reproducible SRIM stopping/range and TRIM calculations on Windows are available at this GitHub repository \href{https://github.com/c-earth/agentic_srim}{https://github.com/c-earth/agentic\_srim.}
 
\section*{Acknowledgements}
DCC acknowledges support from the Jack Kent Cooke Foundation and The Paul and Daisy Soros Fellowships for New Americans. AC acknowledges support from the U.S. National Science Foundation (NSF) Convergence Accelerator Program under Award No. ITE-2345084. MC acknowledges support from the MIT Energy Initiative (MITEI) Fellowship. ML thanks the  support from R.~Wachnik.

\section*{Competing interests}
The authors declare no competing interests. 
 
\bibliographystyle{unsrtnat}
\bibliography{references}

\clearpage
 
\clearpage
\setcounter{page}{1}\renewcommand{\thepage}{S\arabic{page}}
\begingroup
\begin{center}
  {\LARGE\textbf{Supplementary Information for\\[4pt]
  Agentic inverse design of \\ spatially programmed materials}}\par
  \vskip 1.5em
  {\large\lineskip .5em
   \renewenvironment{tabular}[2][]{\begin{center}}{\end{center}}
   \SIsavedauthor\par}
\end{center}
\begin{NoHyper}
\let\thefootnote\relax\footnotetext{$\dagger$ These authors contributed equally to this work.}
\let\thefootnote\relax\footnotetext{$\ast$ Corresponding authors: mingda@mit.edu, denisse@mit.edu}
\end{NoHyper}
\endgroup
\clearpage
\setcounter{figure}{0}\setcounter{table}{0}\setcounter{equation}{0}
\captionsetup{labelsep=space,labelfont=bf,font=small,justification=raggedright,singlelinecheck=false}
\renewcommand{\figurename}{Supplementary Figure}
\renewcommand{\tablename}{Supplementary Table}
\renewcommand{\thefigure}{\arabic{figure}\,}
\renewcommand{\thetable}{\arabic{table}\,}
\renewcommand{\theequation}{S\arabic{equation}}
\renewcommand{\theHfigure}{S\arabic{figure}}     
\renewcommand{\theHtable}{S\arabic{table}}
\renewcommand{\theHequation}{S\arabic{equation}}
\newcounter{snote}
\newcommand{\snote}[1]{\refstepcounter{snote}%
  \section*{Supplementary Note \arabic{snote}: #1}}
\onehalfspacing

\clearpage


\clearpage

\begin{figure}[p]
  \centering
  \includegraphics[width=.5\textwidth]{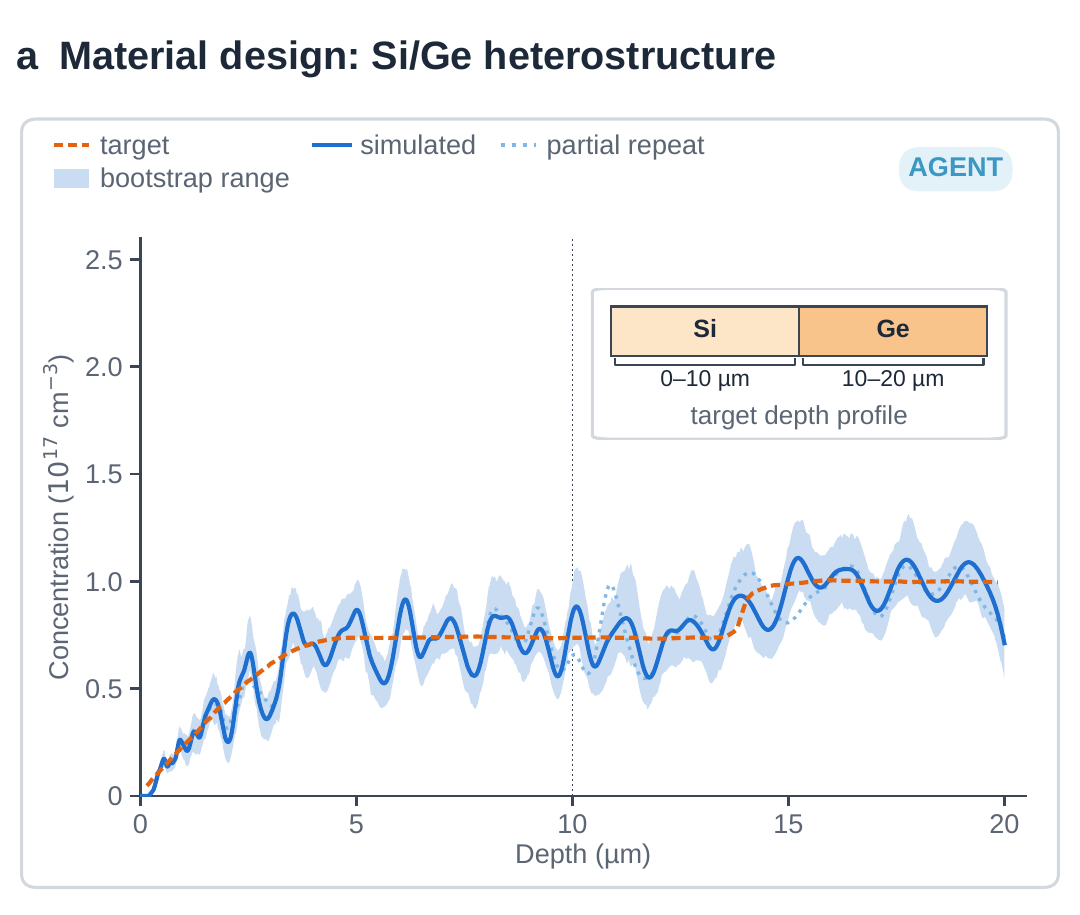}

  \caption{\textbf{The agent designs the two-step concentration profile of Fig.~\ref{fig:gradient} in a Si/Ge bilayer to within the 10\% design criterion, across the interface where the stopping medium changes.}
  \textbf{a},~Simulated Bi profile (blue) compared with the digitized target (orange dashed) in a stack of \SI{10}{\micro\metre} Si on \SI{30}{\micro\metre} Ge, with the beam at normal incidence on the Si. The target is the same hand-drawn sketch and scale as Fig.~\ref{fig:gradient}. It rises from the surface to a plateau near $0.74\times10^{17}$\,cm$^{-3}$ by about \SI{4}{\micro\metre}, runs through the Si/Ge interface (dotted line, \SI{10}{\micro\metre}) and steps up to $1.0\times10^{17}$\,cm$^{-3}$ near \SI{14}{\micro\metre} in the Ge. The inset shows the plotted \SI{0}{}--\SI{20}{\micro\metre} window of the stack.
  The recipe uses 24 energies with nonzero fluencefor a total fluence of $1.50\times10^{14}$\,ions\,cm$^{-2}$. The NRMSE (Eq.~\ref{eq:nrmse}) is 9.31\%. The shading shows a pointwise 95\% bootstrap interval from resampling SRIM ions at fixed dose and kernel bandwidth, and it excludes uncertainty from fit selection and from the model.
  The partial-repeat check (light-blue dotted) replaces the five kernels for which a separate-seed SRIM run exists, together 28.7\% of the total fluence, at fixed fluences and bandwidths, and gives an NRMSE of 9.98\%. The other 19 kernels are unchanged, so this check is not a full independent confirmation of the recipe.
  Profiles are Gaussian kernel density estimates, which blur across the interface, so any concentration discontinuity at \SI{10}{\micro\metre} is unresolved. SRIM treated both layers as amorphous and static with densities of 2.329 (Si) and 5.323\,g\,cm$^{-3}$ (Ge), and the fluences add linearly, which neglects accumulated damage. The Ge displacement and lattice binding energies were set to 15 and 2\,eV without calibration, and the surface binding energy of 3.88\,eV comes from the SRIM ATOMDATA file.}

  \label{supplement Si Ge heterostructure ion implants}
\end{figure}

\begin{figure}[p]
  \centering 

  \includegraphics[width= .9\textwidth]{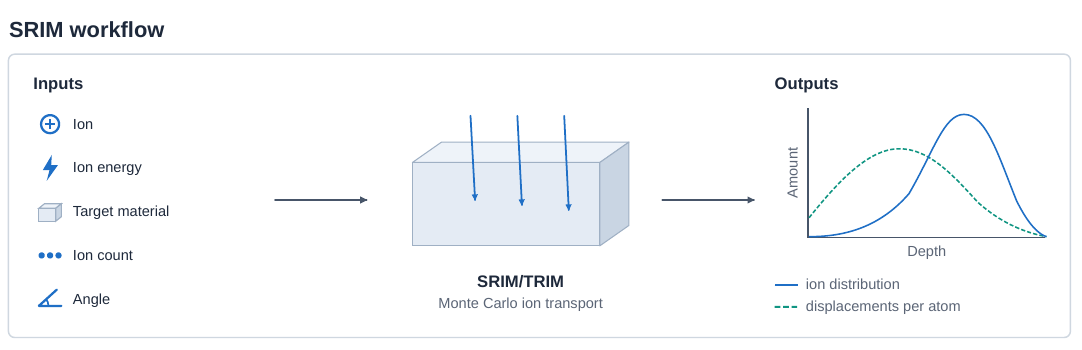}
  \caption{\textbf{A single SRIM call. }Each time the agent calls SRIM through the MCP server (Fig. \ref{fig:architecture}), it sets the ion species, ion energy, target material, ion count and incidence angle. There are additional settings available in the software which are not detailed here. TRIM simulates Monte Carlo ion transport through the target and returns the depth distribution of stopped ions (solid) and the displacements-per-atom damage profile (dashed). Curves are schematic.}
  
  \label{supplement SRIM workflow}
\end{figure}
\begin{figure}[p]
  \centering 

  \includegraphics[width= \textwidth]{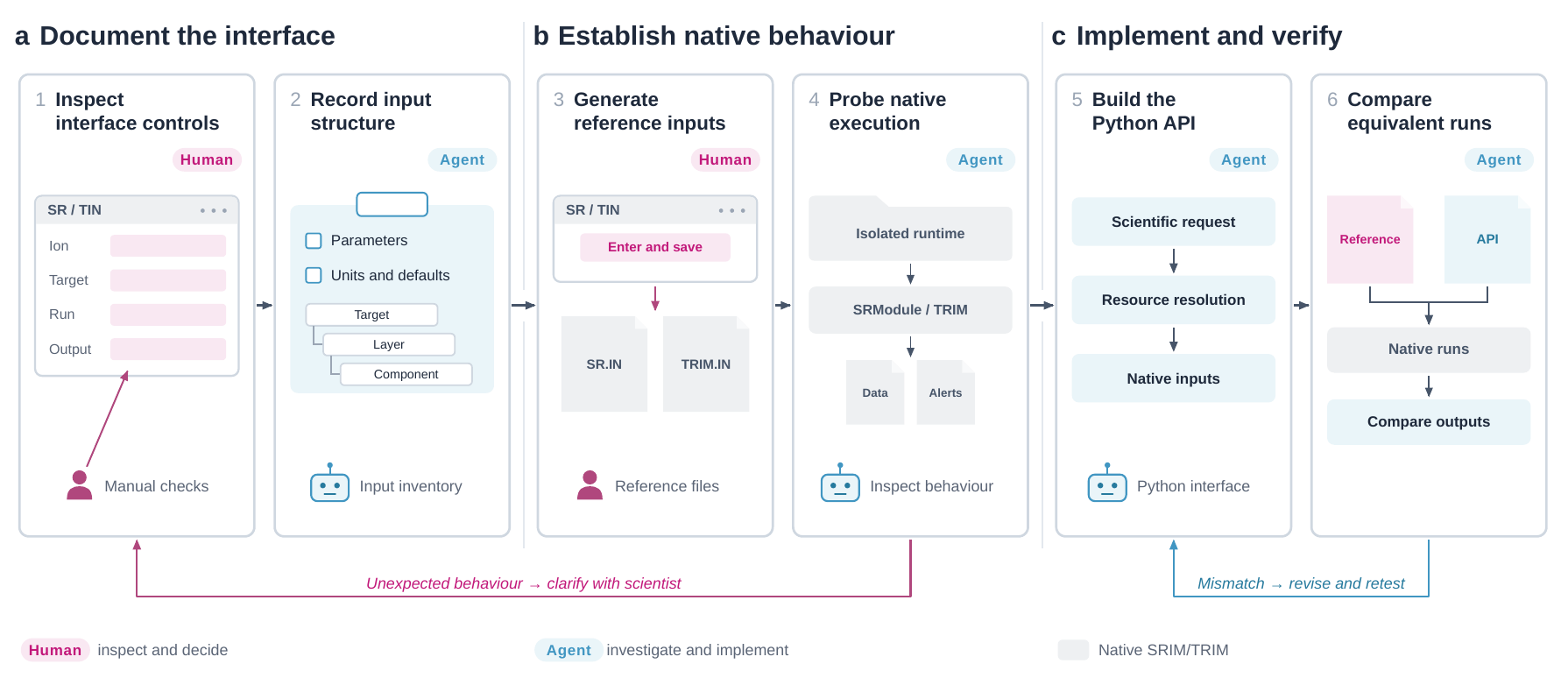}
\caption{\textbf{Agent builds SRIM/TRIM Python API with human input}
\textbf{(a)} Document the interface. The scientist checks each control in the SRIM graphical programs by hand: SR, the stopping and range calculator, and TIN, the TRIM input editor (1). The agent records these controls in an input inventory of parameters, units, defaults and the nested target--layer--component structure of a TRIM target (2).
\textbf{(b)} Establish native behaviour. The scientist enters controlled settings and saves reference input files (\texttt{SR.IN}, \texttt{TRIM.IN}) through the native interface (3). The agent runs these files in isolated copies of the native engines (SRModule and TRIM), finds the files each run needs, and inspects the output data, warnings and saved settings (4). When the native program behaves unexpectedly, the agent asks the scientist to clarify (rose loop).
\textbf{(c)} Implement and verify. The agent writes these rules into a Python API that resolves a scientific request and the selected material and coefficient data into explicit native inputs (5). Reference and API-generated inputs for the same case are run natively and their outputs compared (6). A mismatch sends the agent back to revise the API and retest (blue loop).
Rose marks human actions, blue marks agent and API actions, and grey marks the native program and its files. Interface and file graphics are schematic. The sequence summarizes the recorded development and does not imply that every native option was tested or that SRIM's physics was independently validated.}

  \label{supplement API and GUI}
\end{figure}

\begin{figure}[p]
  \centering
  \includegraphics[width= \textwidth]{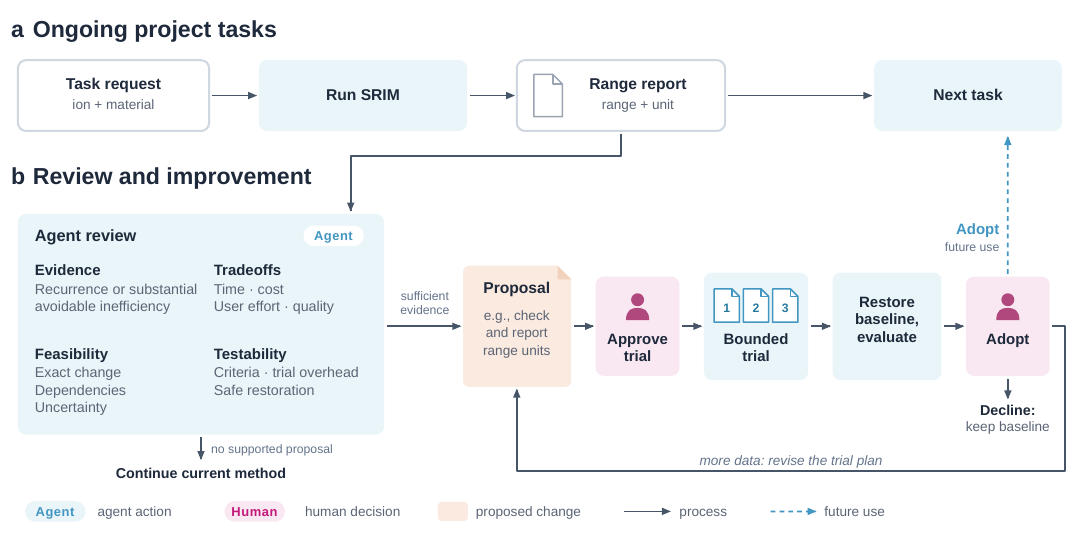}
  \caption{\textbf{Evidence-guided self-improvement built into ordinary project tasks.} (a) The agent completes a scientific task (here, producing an SRIM range report) and carries the result into later work. (b) When a task finishes, the agent reviews it before suggesting any change. The review checks four things. Evidence: a recurring problem across comparable tasks, or one substantial avoidable inefficiency with a clear cause. Tradeoffs: time, cost, user effort and quality. Feasibility: the exact change, its dependencies and its uncertainty. Testability: success criteria, trial overhead and whether the baseline can be safely restored. Without sufficient evidence, the agent continues its current method. With it, the agent writes a specific proposal, which a human must approve before a bounded trial. After the trial, the agent restores the baseline and reports the results. A second, separate human decision then adopts the change for future tasks (dashed arrow), declines it and keeps the baseline, or asks for more data. More data sends the agent back to a revised trial plan and a new approval, so no trial is extended without consent. Blue marks agent actions, pink marks human decisions and orange marks the proposed change. Changes can target the agent’s skills, the workflows and procedures within them, sub-agent definitions and helper scripts, but not the simulator physics. The diagram illustratively shows the project’s self-improvement protocol, which the agent follows through repository instructions, and does not report a measured performance gain.
  }
  \label{supplemental self improvement}
\end{figure}

\begin{figure}[p]
  \centering
  \includegraphics[width= \textwidth]{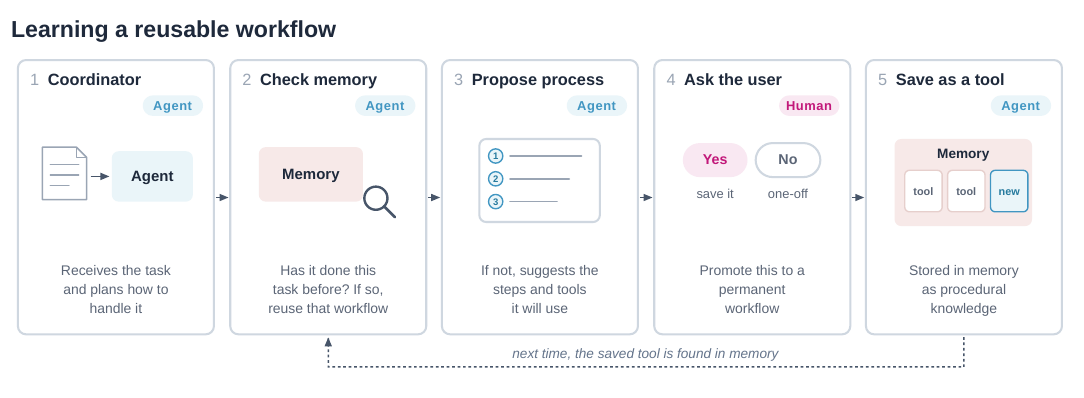}
  \caption{\textbf{Learning a reusable workflow.} (1) The coordinator agent receives a task and plans how to handle it. (2) It checks memory for whether it has done this task before, and if so, reuses the stored workflow. (3) If not, it proposes the steps and tools it will use. (4) The human decides whether to promote this process to a permanent workflow or keep it as a one-off. (5) An approved process is saved as a tool in memory as procedural knowledge, so it is found in step 2 the next time the same task comes up (dashed arrow).}
  \label{supplement procedural memory}
\end{figure}

\begin{figure}[p]
  \centering
  \includegraphics[width =\textwidth]{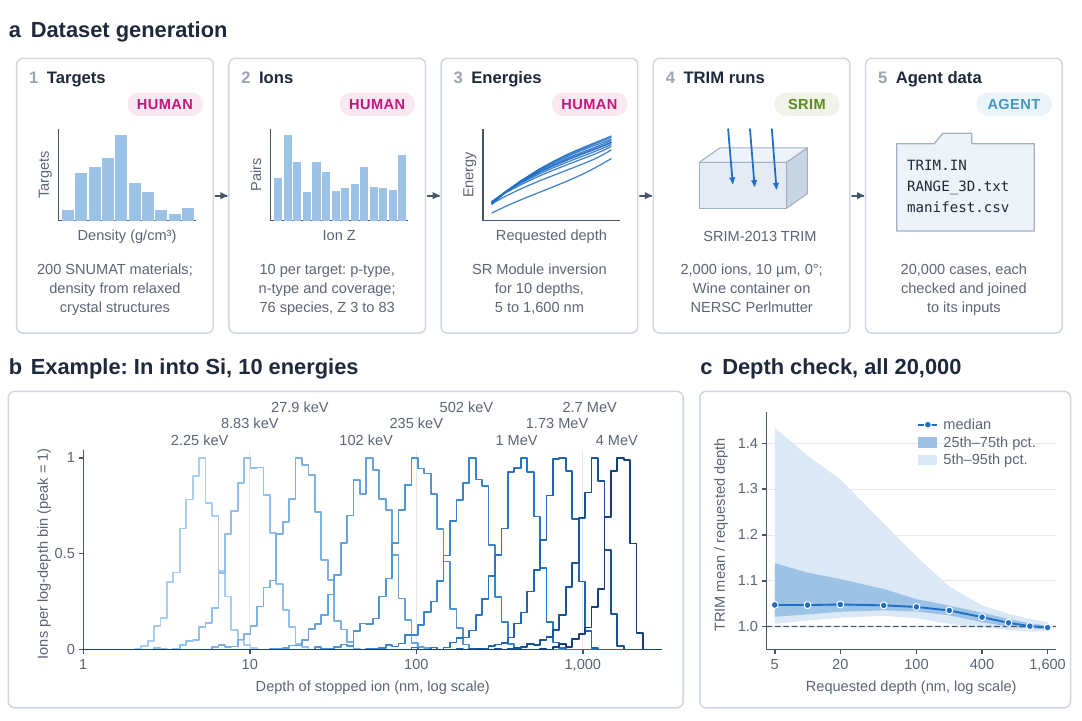}
  \caption{\textbf{Generation of the 20,000-case TRIM dataset available to the agent.} (a) (1) 200 target materials were drawn from the SNUMAT database: 100 chosen for semiconductor application relevance and 100 chosen to broaden coverage of materials space. Each target’s density was computed from its relaxed crystal structure. (2) Each target was paired with 10 ions: recommended p-type and n-type dopants, plus coverage ions chosen to spread the set in atomic number and mass. The set spans 76 species with Z = 3 to 83. (3) For each of the 2,000 target–ion pairs, 10 energies from 0.22 keV to 18.1 MeV were chosen by inverse interpolation of SRIM SR Module range tables. Each energy places the projected range at one of 5, 10, 20, 50, 100, 200, 400, 700, 1,100 or 1,600 nm. (4) Each case was run in TRIM (SRIM-2013.00, Kinchin–Pease quick calculation) with 2,000 ions at normal incidence into a 10 µm target, using a Wine-based container on NERSC Perlmutter. (5) Each stored case contains its input file, the final position of every stopped ion and the run summary, and is linked to its parameters by case ID. An independent check confirmed that all 20,000 cases ran their delivered inputs and that every case accounts for all 2,000 ions. Across the dataset, 2.3\% of ions were backscattered and none were transmitted. Steps 1–3 were designed by the researchers, step 4 was run in SRIM, and step 5 is the data available to the AGENT. Thumbnails show, from left: target densities, ion atomic numbers across target–ion pairs, and selected energy versus requested depth for the 10 ions implanted into Si.
  (b) Example output for In implanted into Si at the 10 selected energies (2.25 keV to 4.0 MeV). Each curve is a histogram of stopped-ion depth, scaled to a peak of 1. Mean depths fall within 3.6
  (c) TRIM mean stopping depth divided by requested depth for all 20,000 cases, excluding backscattered ions: median (line), interquartile range (dark band) and 5th–95th percentiles (light band). The median ratio is 1.05 at 5 nm and 1.00 at 1,600 nm, and 88\% of cases are within 10\% of the requested depth. The largest deviations are for light ions at shallow depths, where up to 41\% of ions backscatter.
}
  \label{supplement 20000 database}
\end{figure}

\begin{figure}
    \centering
    \includegraphics[width=\textwidth]{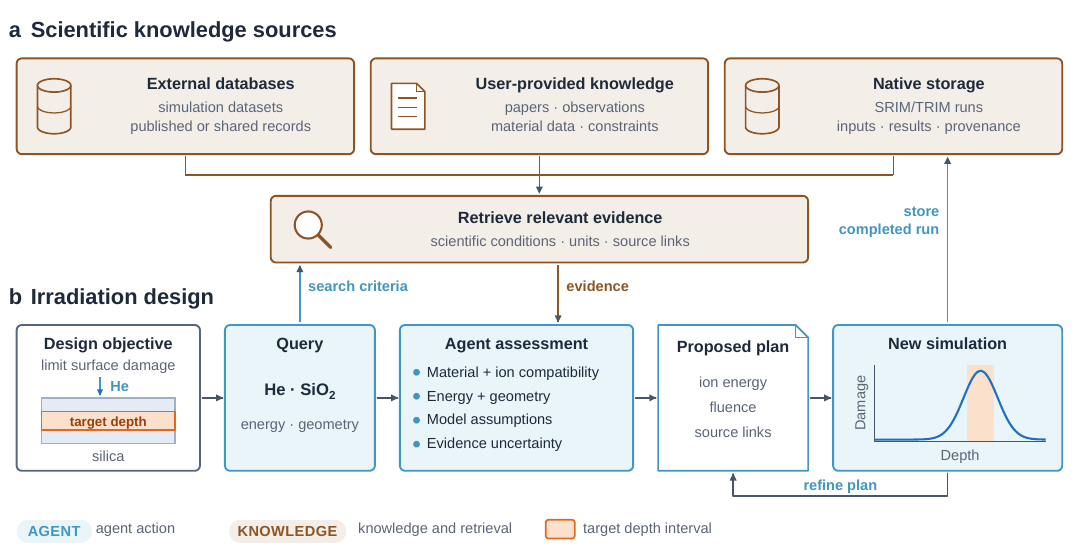}
    \caption{(a) Three knowledge sources feed a retrieval layer that keeps scientific conditions, units and source links. External databases hold published or shared simulation datasets. User-provided knowledge includes papers, experimental observations, material data and constraints. Native storage holds the inputs, results and provenance of the project’s own SRIM (SR Module and TRIM) runs. (b) A design objective is translated into search criteria. The example objective is helium irradiation of silica to concentrate damage in a target depth interval while limiting near-surface damage. The agent assesses the retrieved evidence for material and ion compatibility, energy and geometry, model assumptions and evidence uncertainty. It then proposes irradiation conditions (ion energy and fluence) with supporting source links, tests the plan in a new simulation, and refines the plan if the simulated damage profile misses the target interval. Each completed run is stored with its inputs, results and provenance in native storage, adding to the evidence for future searches. Blue marks agent actions, brown marks knowledge and retrieval, and orange marks the target depth interval. The architecture and damage profile are schematic, and the profile is illustrative rather than simulated. In the current implementation, retrieval is exact structured search over the verified saved inputs of registered native SRIM runs. External databases, user-provided knowledge, search over output values and the full design cycle are the planned architecture and are not yet implemented.}
    \label{supplement RAG}
\end{figure}  

\begin{figure}[p]
  \centering
 \includegraphics[width= \textwidth]{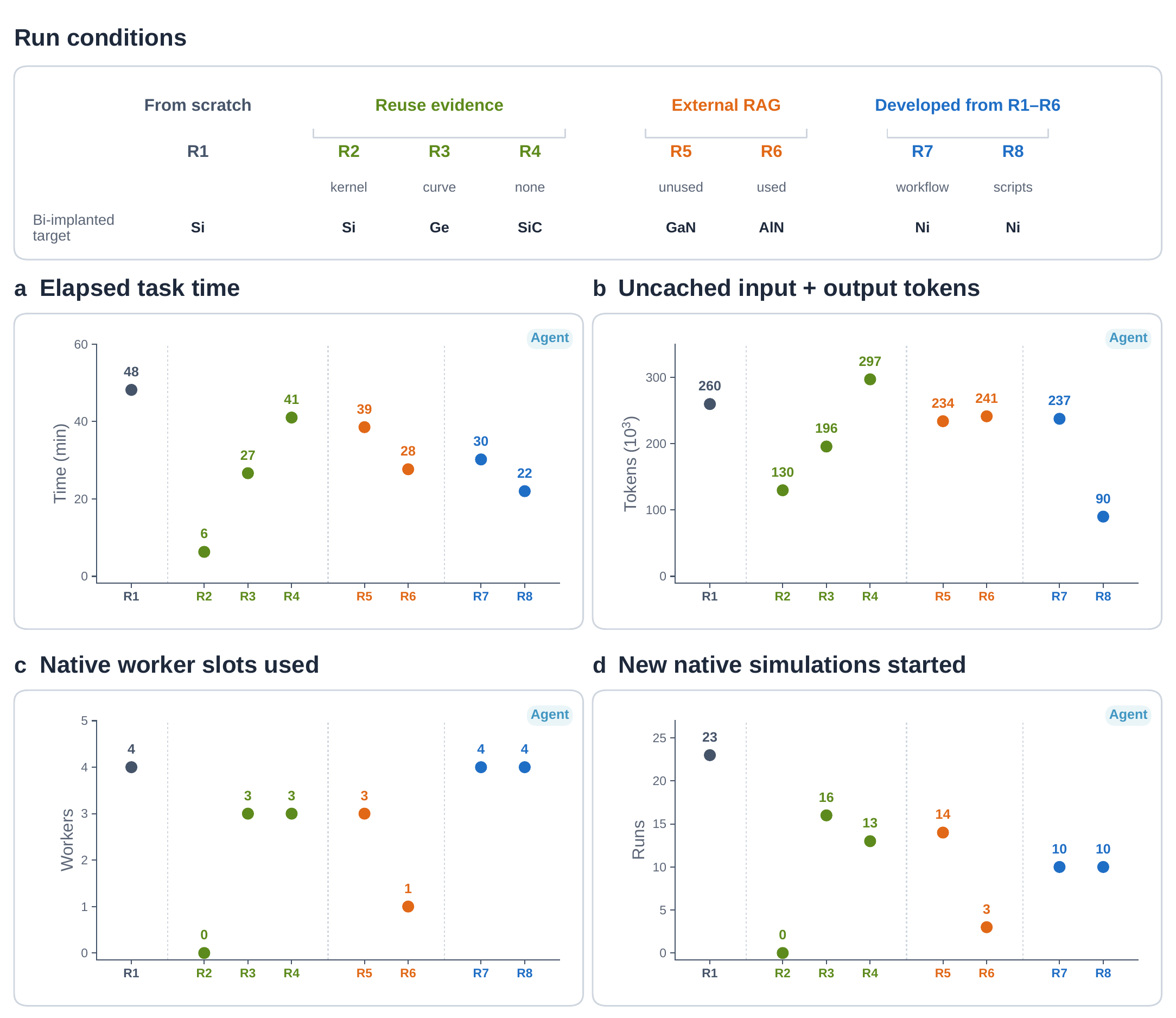}
  \caption{\textbf{Reusing earlier evidence lowers the time, tokens and simulations a design needs.}
  Top, conditions of the eight runs, each designing a Bi implantation profile in the target listed. R1 started from scratch (grey). R2--R4 reused evidence (green): a SRIM kernel, a target curve or none applicable. R5 and R6 used external retrieval-augmented generation (RAG; orange), with the retrieved numerical profiles unused or used. R7 ran the revised workflow developed from R1--R6 (blue). R8 is an estimate, not a measured run, for the R7 task carried out with reusable scripts.
  \textbf{a},~Elapsed task time.
  \textbf{b},~Uncached input plus output tokens. Cached input is part of the input and reasoning tokens are part of the output, so neither is counted twice.
  \textbf{c},~Number of SRIM worker slots used in parallel.
  \textbf{d},~Number of new SRIM simulations started, including failed and continued runs.
  Values above each point are rounded to whole units. R1--R7 are measured from the original runs, with token counts checked against ccusage 20.0.41. R8 assumes that a validated script already exists, keeps R7's 10 simulations, four workers, checks and review steps, and removes only the bookkeeping time outside SRIM and timed tool calls. The upper of two scenarios (20.0--\SI{21.6}{min}; 57--89 thousand tokens), rounded up, gives \SI{22}{min} and 90 thousand tokens. It excludes the one-time cost of writing the script and has not been confirmed by a scripted run. Final NRMSE values are 4.83, 4.26, 5.68, 4.29, 4.58, 6.63 and 9.87\% for R1--R7. Runs differ in material, target and history, so the panels do not isolate the effect of each condition.}
  \label{supplement fig:evaluation}
\end{figure}

\end{document}